\documentclass{IEEEtran}
\usepackage[T1]{fontenc}
\usepackage{graphicx}
\usepackage{mathtools}
\usepackage{amssymb}
\usepackage{amsthm}
\usepackage{thmtools}
\usepackage{xcolor}
\usepackage{nameref}
\usepackage[colorlinks=true, linkcolor=blue, citecolor=blue, urlcolor=blue]{hyperref}
\usepackage{cite}
\usepackage{textcomp}
\usepackage{amsmath}
\usepackage{autobreak}
\usepackage{amsfonts}
\usepackage{bm}
\usepackage{bbm}
\usepackage{algorithm}  
\usepackage{algorithmicx}  
\usepackage{algpseudocode}
\usepackage{stfloats}
\usepackage{cuted}
\usepackage{color}
\usepackage{booktabs} 
\usepackage{soul}
\usepackage{arydshln}
\usepackage{cases}
\usepackage{comment}
\usepackage{subcaption}
\title{Movable Antenna Enhanced MIMO Communications with Spatial Modulation}
\author{Kaihe Wang,
	\IEEEmembership{Graduate Student Member, IEEE}, Ran Yang, \IEEEmembership{Graduate Student Member, IEEE},\\
     Lipeng Zhu, \IEEEmembership{Member, IEEE}, Rongyan Xi, Yue Xiu, \IEEEmembership{Member, IEEE}, Zhongpei Zhang
	\thanks{Kaihe Wang, Ran Yang, Yue Xiu, and Zhongpei Zhang are with the University of Electronic Science and Technology of China (UESTC), Chengdu 611731, China (E-mail: khewang@yeah.net, zhangzp@uestc.edu.cn).
    
    Lipeng Zhu is with the School of Interdisciplinary Science, Beijing Institute of Technology, Beijing 100081, China (E-mail: lipzhu@outlook.com).
    
    Rongyan Xi is with the Future Research Laboratory, China Mobile Research Institute, Beijing 100053, China (E-mail: xirongyan@chinamobile.com).}
}

\begin{document}
	\maketitle
	\begin{abstract}
		Movable antenna (MA) has demonstrated great potential in enhancing wireless communication performance. In this paper, we investigate an MA-enabled multiple-input multiple-output (MIMO) communication system with spatial modulation (SM), which improves communication performance by utilizing flexible MA placement while reducing the cost of RF chains. To this end, we propose a joint transceiver design framework aimed at minimizing the bit error rate (BER) based on the maximum minimum distance (MMD) criterion. To address the intractable problem, we develop an efficient iterative algorithm based on alternating optimization (AO) and successive convex approximation (SCA) techniques. Simulation results demonstrate that the proposed algorithm achieves rapid convergence performance and significantly outperforms the existing benchmark schemes.
	\end{abstract}
	
	\section{Introduction}
    \IEEEPARstart{T}{he} International Mobile Telecommunications (IMT) 2030 vision identifies ultra-massive multiple-input multiple-output (MIMO) as an indispensable technology for air interface. However, scaling antenna arrays to ultra-massive dimensions significantly increases hardware costs and power consumption due to the need for configuring a large number of dedicated radio frequency (RF) chains. To address this issue while preserving high spectral efficiency, spatial modulation (SM) has been integrated into the massive MIMO framework\cite{mesleh2008spatial}. 
    
    By embedding information into the index of activated  antennas, SM effectively strikes a balance between high spectral efficiency (SE) and low hardware complexity. Moreover, recent research progress has enabled SM to be applied in a variety of emerging wireless communication scenarios, including orthogonal time frequency space modulation\cite{zhang2021spatial}, intelligent space-air-ground communication systems, and reconfigurable intelligent surface-assisted communications\cite{zhu2025transmissive}. Although the aforementioned works have demonstrated the efficacy of SM, they are mainly based on a fixed-position antenna (FPA) transceiver prototypes. By restricting transceivers to static locations, FPAs fail to fully exploit the fine-grained channel variations in the continuous spatial field. As such, the fixed geometric configurations of FPAs inevitably result in channel-gain loss, particularly when optimal channel conditions deviate from the static array topology\cite{zhu2025tutorial}. These limitations constitute a fundamental bottleneck of SM performance, thereby impeding further breakthroughs in system reliability and capacity.
    
    Recently, movable antennas (MAs), also known as fluid antennas (FAs)\cite{wong2021fluid, zhu2024movable}, have emerged as a transformative solution to address the limitations of FPAs by properly adjusting antenna placement. Owing to the significant flexibility and performance gains offered by MAs, researchers have recently expanded their application to the field of index modulation (IM), effectively boosting spectral efficiency without increasing hardware complexity\cite{zhu2024index}. In fact, the primary motivation for introducing IM based on SM is to extend advantages of SM in MIMO systems, such as high energy efficiency, interference resistance, and low complexity, to broadband multi-carrier systems that do not require multiple antennas. Consequently, to fully leverage the inherent advantages of SM-MIMO in complex scenarios such as millimeter-wave communication and Internet of Things, we proposed an SM-MIMO communication system with MAs. It is based on dynamic optimization of the antenna position according to channel conditions, thus significantly enhancing spatial diversity and robustness. Ultimately, this flexibility effectively mitigates the impact of complex channel variations, leading to a substantial improvement in the system’s bit error rate (BER) performance.
    
    Motivated by the above considerations, we study an MA-enabled SM-MIMO system in this paper. To minimize the BER, we formulate an optimization problem by jointly optimizing the transmit beamforming and antenna placement based on the minimum distance (MMD) criterion. To address the intractable optimization problem, we develop an efficient alternating optimization (AO)-based algorithm to iteratively update the precoder and antenna positions, subject to the constraints of maximum transmit power, antenna moving regions, and minimum inter-MA distance. We show that the proposed algorithm reaches a fast convergence rate and exhibits a significantly lower BER compared to conventional SM-MIMO systems.
	\begin{figure}[h]
		\centering
		\includegraphics[width=3.4in]{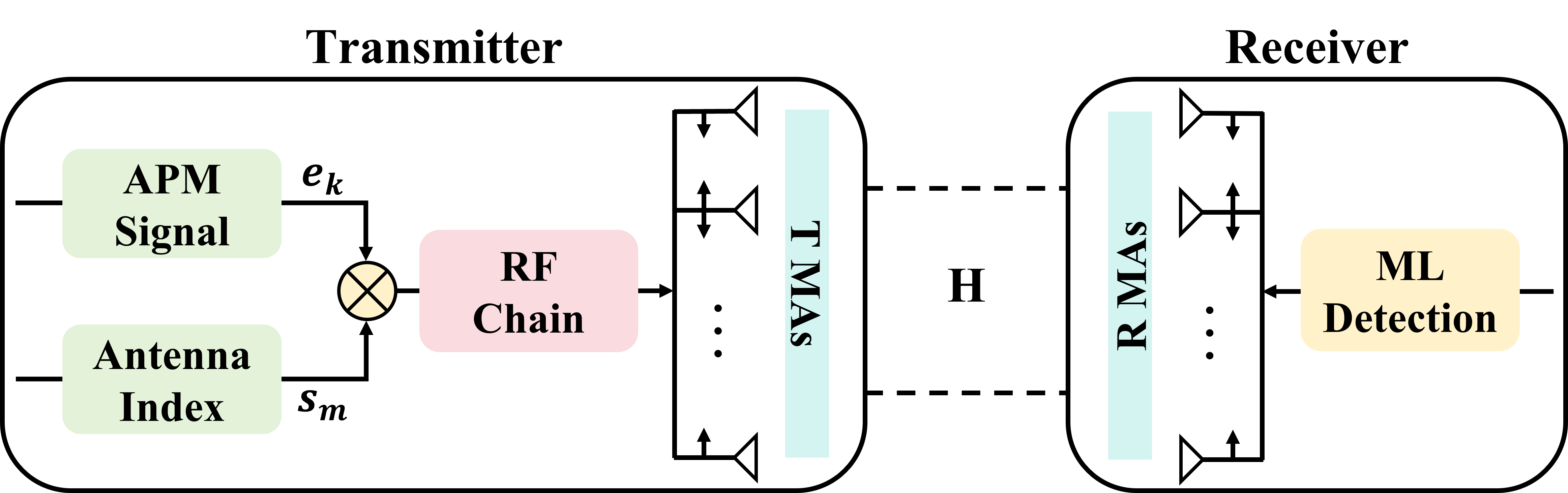}
		\caption{System model of the SM-MIMO system with MAs.}
		\label{system_module}
	\end{figure}
	\section{System Model and Problem Formulation}
    \subsection{Channel Model}
	Consider an SM-MIMO system where both transmitter and receiver are equipped with linear arrays containing \( T \) and \( R \) MAs, respectively, as shown in Fig.\ref{system_module}. Since the size of its transmit/receive region is much smaller than the propagation distance of the signal, the plane wave model can be used to construct the field-response channel model between transmitter and receiver under the far-field assumption\cite{zhu2024modeling}.
	
	Denoting the transmitter and receiver reference points as \(O_t\) and \(O_r\), the transmitting and receiving antennas are moved over one-dimensional regions \([0, A_t]\) and \([0, A_r]\), respectively. The positions of the transmit MAs are denoted as \( \boldsymbol u = [u_1, \ldots, u_T] \in [0, A_t] \), where \(u_t\) indicates the position of the \( t \)-th antenna. Define the angles of departure (AoDs) and the angles of arrival (AoAs) as \( \theta_i (i = 1, 2, \ldots, L_t) \) and \( \phi_j (j = 1, 2, \ldots, L_r) \), respectively. Then, the transmit field-response vector (FRV) is given by
	\begin{align*}
		\boldsymbol{g}(u_t) &= \left[ e^{j \frac{2\pi}{\lambda} u_t \cos \theta_1}, \ldots, e^{j \frac{2\pi}{\lambda} u_t \cos \theta_{L_t}} \right]^T \in \mathbb{C}^{L_t \times 1}, \tag{1}
	\end{align*}
	where \(u_t \cos \theta_{l_t}\) represents the propagation distance difference of the \(L_t\)-th transmit path between the \(t\)-th antenna of the transmitter and \(O_t\).
	Therefore, the transmit field-response matrix (FRM) is given by
	\begin{align*}
		\mathbf{G}(\boldsymbol u) &= [\boldsymbol{g}(u_1), \ldots, \boldsymbol{g}(u_T)] \in \mathbb{C}^{L_t \times T}, \tag{2}
	\end{align*}
	Similarly, the receive FRV and FRM are given by
	\begin{align*}
		\boldsymbol{f}(v_r) &= \left[ e^{j \frac{2\pi}{\lambda} v_r \cos \phi_1}, \ldots, e^{j \frac{2\pi}{\lambda} v_r \cos \phi_{L_r}} \right]^T \in \mathbb{C}^{L_r \times 1}, \tag{3}\\
		\mathbf{F}(\boldsymbol v) &= [\boldsymbol{f}(v_1), \ldots, \boldsymbol{f}(v_R)] \in \mathbb{C}^{L_r \times R}, \tag{4}
	\end{align*}
	where \(v_r \cos \phi_{l_r}\) represents the propagation distance difference of the \(L_r\)-th receive path between the \(r\)-th antenna of receiver and \(O_r\).
	
	Define the path response matrix (PRM) \( \mathbf{\Sigma} \in \mathbb{C}^{L_r \times L_t} \) containing the response coefficients of all paths from \(O_t\) to \(O_r\). Thus, the channel matrix from transmitter to receiver can be expressed as
	\begin{align*}
	&\mathbf{H}(\boldsymbol u, \boldsymbol v) = \mathbf{F}(\boldsymbol{v})^{H} \mathbf{\Sigma} \mathbf{G}(\boldsymbol{u}) \in \mathbb C^{R\times T}. \tag{5}
	\end{align*}
	In the subsequent discussion, assuming that the path characteristics remain stable in the short term, \(\mathbf{\Sigma}\) can be regarded as a constant. When both transmitter and receiver are equipped with MAs, \( \mathbf{F} \) is a function of \( \boldsymbol v \) and \( \mathbf{G} \) is a function of \( \boldsymbol u \).
	\subsection{Signal Model}
	The SM signal model introduces a spatial dimension to the amplitude and phase modulation (APM) signal by utilizing the additional space of the antenna indices. The indicator vector \( \mathbf{e}_k = [0, 0, \ldots, 1, \ldots, 0, 0]^T \in \mathbb C^{T\times 1} \), with only the \(k\)-th element is 1, corresponds to the constellation point in the spatial constellation space as the activated antenna indices. For \( M \)-order modulation, the spatial and signal constellations jointly transmit \(\log_2 T + \log_2 M\) bits per channel use (bpcu)\cite{lee2015generalized}. Thus, the transmit codebook \(\mathcal{S}\) with cardinality \(TM\) can be constructed through the Cartesian product of spatial constellation points and signal constellation points. The transmitted SM symbol \(\mathbf{x} = \mathbf{e}_k s_m \in \mathbb C^{T\times 1}\) is selected from \(\mathcal{S}\), where \( k = 1, 2, \ldots, T \) and \( m = 0, 1, \ldots, M-1 \).
	
	To isolate the effective channel vectors between transmit antennas and enhance the BER performance, a diagonal matrix \(\mathbf{W} = \text{diag} \{\boldsymbol w\} \in \mathbb C^{T\times T}\) is used to precode \(\mathbf{x}\), where \(\boldsymbol w = [w_1, \ldots, w_T]^T\) represents the precode weights. Based on this, the received signal can be given by
	\begin{align*}
		&\mathbf{y} = \mathbf{H}\mathbf{W}\mathbf{x} + \mathbf{n}, \tag{6}
	\end{align*}
	where \(\mathbf{H} \in \mathbb C^{R\times T}\) is the flat-fading MIMO channel matrix, and \(\mathbf{n} \sim \mathcal{CN}(\mathbf{0}, \sigma^2 \mathbf{I}_R)\) is the additive white Gaussian noise (AWGN).
    
    To achieve the optimal error performance, SM signal demodulation traverses all \( \mathbf{x} \in \mathcal{S} \) based on the maximum likelihood (ML) detection algorithm as
	\begin{align*}
		\hat{\mathbf{x}}_{\text{ML}} &= \arg \min \limits_{\mathbf{x} \in \mathcal{S}}
		\|\mathbf{y} - \mathbf{H}\mathbf{W}\mathbf{x}\|^2. \tag{7}\label{7}
	\end{align*}
	
	Since (\ref{7}) lacks of a closed-form solution, the upper bound of BER for MIMO channel can be approximated by the total pairwise error probability (Total PEP)\cite{yang2014design}. Assume that the codebook symbols are transmitted with equal probability. As signal-to-noise ratio (SNR) increases, the minimum distance determines the probability of error detection. Consequently, the Total PEP can be further simplified to
	\begin{align*}
		&P_{e}^{o} \approx \iota \cdot Q \left(\sqrt{\frac{1}{2\sigma^{2}} d_{\min}(\boldsymbol w, \boldsymbol u, \boldsymbol v)}\right), \tag{8}
	\end{align*}
	where \(\iota\) is the number of signal pairs in the codebook \(\mathcal{S}\), and \(d_{\min}(\boldsymbol w, \boldsymbol u, \boldsymbol v)\) represents the minimum squared Euclidean distance among all signal pairs defined as
	\begin{align*}
		&d_{\min}(\boldsymbol w, \boldsymbol u, \boldsymbol v) = \min \limits_{\mathbf{x}_i, \mathbf{x}_j \in \mathcal{S}, i\neq j} \|\mathbf{H}\mathbf{W}(\mathbf{x}_i-\mathbf{x}_j)\|^2. \tag{9}\label{9}
	\end{align*}
	\subsection{Problem Formulation}
    By leveraging the monotonically decreasing property of the Q-function, the system optimization objective can be transformed into the MMD problem. This goal is achieved by jointly optimizing the transmit precoder \(\mathbf W\), the transmit antenna positions \(\boldsymbol u\), and the receive antenna postions \(\boldsymbol v\). By introducing the auxiliary variable \(\eta \triangleq d_{\min}(\boldsymbol w, \boldsymbol u, \boldsymbol v)\), the optimization problem can be formulated as
	\begin{align*}
		&\max \limits_{\eta, \boldsymbol w, \boldsymbol u, \boldsymbol v}\quad \eta \tag{10a}\\
		&{\mathrm{s.t.}}\quad \|\mathbf{H}(\boldsymbol u, \boldsymbol v) \text{diag} \{\boldsymbol w\} (\mathbf{x}_i-\mathbf{x}_j)\|^2 \geq \eta, \tag{10b}\label{10b}\\
		&\quad \hphantom{s.t. } \|\boldsymbol w\|^2 \leq P_T, \tag{10c}\label{10c}\\
		&\quad \hphantom{s.t. } \boldsymbol u \in [0, A_t], \ \boldsymbol v \in [0, A_r], \tag{10d}\label{10d}\\
		&\quad \hphantom{s.t. } |u_a - u_b|\geq D, \ u_a, u_b\in \boldsymbol u, \ \forall a\neq b, \tag{10e}\label{10e}\\
		&\quad \hphantom{s.t. } |v_a - v_b|\geq D, \ v_a, v_b\in \boldsymbol v, \ \forall a\neq b. \tag{10f}\label{10f}
	\end{align*}
    In (\ref{10c}), \(P_T\) indicates the maximum power budget for the transmitter. (\ref{10d}) limits the antenna movement range, and (\ref{10e}) and (\ref{10f}) represent the minimum constraints on the antenna spacing to avoid coupling effects. Since \(\boldsymbol w\), \(\boldsymbol u\) and \(\boldsymbol v\) are intricately coupled in the non-convex constraints (\ref{10b}), the problem is challenging to solve. Thus, we devise an efficient algorithm to solve this problem in the next section.
    \section{The Proposed Solution}
	In this section, an efficient AO-based algorithm is proposed to solve the considered MMD problem. The algorithm decouples the optimization variables \(\boldsymbol w\), \(\boldsymbol u\) and \(\boldsymbol v\) and decomposes them into three sub-problems to be optimized respectively. For the \(n\)-th iteration, \(\boldsymbol w\) is first updated with fixed \(\boldsymbol u\) and \(\boldsymbol v\). Subsequently, \(\boldsymbol u\) and \(\boldsymbol v\) are updated sequentially using a block coordinate descent (BCD) approach.
	\subsection{Optimizing Transmit Beamforming}
    \begin{figure*}[hb]
		\centering
		\vspace*{0pt}
		\hrulefill
		\vspace*{0pt} 
		\begin{align*}
			\|\mathbf{H}(\boldsymbol u, \boldsymbol v)\mathbf{W}_{\mathrm{D}}(\mathbf{x}_{m}^{k}-\hat{\mathbf{x}}_{\hat{m}}^{\hat{k}})\|^2 
			&= \|\mathbf{B}\mathbf{G}(\boldsymbol u)\mathbf{c}\|^2 = (c_k \mathbf{B} \boldsymbol{g}(u_k) + \mathbf{\Lambda})^H (c_k \mathbf{B} \boldsymbol{g}(u_k) + \mathbf{\Lambda}) \\
			&= |c_k|^2 \boldsymbol{g}(u_k)^H \mathbf{B}^H \mathbf{B} \boldsymbol{g}(u_k) + 2\Re\left\{ (c_k \mathbf{B} \boldsymbol{g}(u_k))^H \boldsymbol{\Lambda} \right\} + \|\mathbf{\Lambda}\|^2 \\
			&= 2|c_k|^2 \sum_{i=1}^{L_t-1} \sum_{j=i+1}^{L_t} \left| \left[ \mathbf{B}^H \mathbf{B} \right]_{i,j} \right| \cos \left(\alpha(u_k)\right) + 2|c_k| \sum_{r=1}^{R} \sum_{i=1}^{L_t} \left| \left[ \mathbf{B} \right]_{r,i} \right| |\mathbf{\Lambda}_r| \cos \left(\beta(u_k)\right) \\
			& \quad + |c_k|^2 \sum_{i=1}^{L_t} \left[ \mathbf{B}^H \mathbf{B} \right]_{i,i} + \sum_{r=1}^{R} |\mathbf{\Lambda}_r|^2\triangleq \; y(u_k). \tag{14}\label{14}
		\end{align*}
	\end{figure*}
	With fixed antenna positions, the optimization problem can be simplified as
	\begin{align*}
		&\max \limits_{\eta, \boldsymbol w}\quad \eta \tag{11a} \\
		&{\mathrm{s.t.}}\quad \|\mathbf{H} \text{diag} \{\boldsymbol w\} (\mathbf{x}_i-\mathbf{x}_j)\|^2 \geq \eta, \tag{11b}\label{11b}\\
		&\quad \hphantom{s.t. } \text{(\ref{10c})}.
	\end{align*}
	However, the constraint (\ref{11b}) is still non-convex due to the quadratic form of \(\boldsymbol w\). To solve this problem, SCA technique is adopted to obtain the solution by iteratively approximating this non-convex constraint\cite{gao2024joint}. Specifically, the left-hand-side of (\ref{11b}) is replaced by its global under-estimator, which is derived from the first-order Taylor expansion of the current (\(n\)-th) iteration point \(\boldsymbol w^n\). For convenience, we define \(\mathbf{X} \triangleq \text{diag} \left\{(\mathbf{x}_i-\mathbf{x}_j)\right\}\) and \(\mathbf{A} \triangleq \mathbf{X}^H \mathbf{H}^H \mathbf{H} \mathbf{X}\), so that the constraint (\ref{11b}) can be approximated as
	\begin{align*}
		& 2\Re\left\{ (\boldsymbol w^n)^H \mathbf{A} \boldsymbol w \right\} - (\boldsymbol w^n)^H \mathbf{A} \boldsymbol w^n \geq \eta. \tag{12}\label{12}
	\end{align*}
	
	At this point, the original non-convex problem is transformed into a convex quadratically constrained program (QCP) problem, which can be solved directly by using existing tools such as CVX\cite{grant2008cvx}.
	\subsection{Optimizing Antenna Position}
	In this section, we propose a BCD-based optimization on antenna potisions. With other optimization variables being fixed, the optimization problem with respect to \(u_k\) can be given by
	\begin{align*}
		&\max \limits_{\eta,u_k}\quad \eta \tag{13a}\\
		&{\mathrm{s.t.}}\quad \|\mathbf{H}(\boldsymbol u, \boldsymbol v)\mathbf{W}(\mathbf{x}_i-\mathbf{x}_j)\|^2 \geq \eta, \tag{13b}\label{13b}\\
		&\quad \hphantom{s.t. } u_k \in [0, A_t], \tag{13c}\label{13c}\\
		&\quad \hphantom{s.t. }
		|u_k - u_c|\geq D, \ \forall c\in T, c\neq k. \tag{13d}\label{13d}
	\end{align*}
	
	The optimization of variable \(u_k\) requires deriving its explicit expression from the constraint (\ref{13b}). Define \(\mathbf{B} \triangleq \mathbf{F}(\boldsymbol v)^{H} \mathbf{\Sigma}, \mathbf{c} \triangleq \mathbf{W}(\mathbf{x}_i-\mathbf{x}_j) \) and expand constraint (\ref{13b}) in (\ref{14}) at the bottom of this page, where \( \boldsymbol \xi \triangleq \sum_{t=1, t \neq k}^{T} c_t \mathbf{B} \boldsymbol{g}(u_t) \), \( \alpha(u_k) \triangleq \frac{2\pi}{\lambda} u_k (\cos \theta_j - \cos \theta_i) + \arg \left( \left[ \mathbf{B}^H \mathbf{B} \right]_{i,j} \right) \), and \( \beta(u_k) \triangleq -\frac{2\pi}{\lambda} u_k \cos \theta_i - \arg \left( \left[ \mathbf{B} \right]_{r,i} \right) - \arg(c_k) + \arg(\xi_r) \). Since \(y(u_k)\) is neither convex nor concave, it is necessary to construct a concave lower-bound surrogate function \(y_\mathrm{lb}^n(u_k)\) through second-order Taylor expansion in order to transform the non-convex problem into a convex problem
	\begin{align*}
		y(u_k) &\geq y(u_k^n) + \nabla y(u_k^n)(u_k - u_k^n) - \frac{\epsilon_k}{2} (u_k - u_k^n)^2 \\
		&\triangleq \; y_\mathrm{lb}^n(u_k), \tag{15}\label{15}
	\end{align*}
	where \(u_k^n\) denotes the value obtained during the \(n\)-th iteration of optimizing a single antenna \(u_k\), and \(\epsilon_k\) is a positive real number satisfying \(\epsilon_k\mathbf {I} - \nabla^{2}y(u_k) \succeq \boldsymbol 0\)\cite{gao2024joint}. The expansion of \(\nabla y(u_k)\) and \(\nabla^{2} y(u_k)\) is given in (\ref{16}) and (\ref{17}) at the bottom of the next page, respectively.
	
	Therefore, the problem of optimizing \(u_k\) can now be approximated as
	\begin{align*}
		&\max \limits_{\eta,u_k}\quad \eta \tag{18a}\\
		&{\mathrm{s.t.}}\quad y_\mathrm{lb}^n(u_k) \geq \eta, \tag{18b}\\
		&\quad \hphantom{s.t. } \text{(\ref{13c})},\text{(\ref{13d})}.	
	\end{align*}
	At this stage, the optimization problem is convex and can be solved by off-the-shelf solvers. Similarly, the optimization of receiver antennas can be completed efficiently in the same way, as detailed in the appendix. The overall optimization procedure is summarized in Algorithm \ref{Al}.
	\subsection{Complexity Analysis of Algorithm}
	Based on the above solution process, Algorithm \ref{Al} provides an overview of the proposed AO-based algorithm for MMD problem. In each sub-problem, the optimization variables reach local optimum solution, and Algorithm \ref{Al} converges. Algorithm \ref{Al} ultimately converges to a local optimum. The complexity analysis of Algorithm \ref{Al} is as follows. Firstly, the complexity of updating \(\boldsymbol w\) at step 4 is considered to be \(\mathcal{O}\left(T^{3.5}\right)\). Next, the computational cost of updating \(\boldsymbol u\) and \(\boldsymbol v\) are \(\mathcal{O}\left(T^{4.5}\right)\) and \(\mathcal{O}\left(TR^{3.5}\right)\) respectively. Therefore, the total complexity of Algorithm \ref{Al} is \(\mathcal{O}\left(\ln\frac{1}{\kappa} \left(T^{4.5}+TR^{3.5}\right)\right)\)\cite{gao2024joint}.
	\begin{algorithm}[H]
		\caption{Proposed AO-Based Algorithm for MMD} \label{Al}
		\begin{algorithmic}[1]
			\State Set iteration counter \(n = 0\), convergence threshold \(\kappa\).
			\State Initialize \(\boldsymbol w^0, \boldsymbol u^0, \boldsymbol v^0\).
			\Repeat
			\State Update \(\boldsymbol w^{n+1}\) for given \(\boldsymbol u^n\) and \(\boldsymbol v^n\).
			\State Update \(\boldsymbol u^{n+1}\) for given \(\boldsymbol w^n\) and \(\boldsymbol v^n\).
			\For{\(u_k^n = u_1^n\) to \(u_T^n\)}
			\State Update \(u_k^{n+1}\) for given \(\boldsymbol w^n\), \(\boldsymbol u^n\) expect \(u_k^n\) and \(\boldsymbol v^n\).
			\EndFor
			\State Update \(\boldsymbol v^{n+1}\) for given \(\boldsymbol w^n\) and \(\boldsymbol u^n\).
			\For{\(v_k^n = v_1^n\) to \(v_R^n\)}
			\State Update \(v_k^{n+1}\) for given \(\boldsymbol w^n\), \(\boldsymbol u^n\) and \(\boldsymbol v^n\) expect \(v_k^n\).
			\EndFor
			\State \(n \leftarrow n + 1\)
			\Until{The objective function value converges.}
			\Ensure Optimal solution \(\boldsymbol w^*, \boldsymbol u^*, \boldsymbol v^*\)
		\end{algorithmic}
	\end{algorithm}
	\section{Simulation Results}
    \begin{figure*}[hb]
		\centering
		\vspace*{0pt}
		\hrulefill
		\vspace*{0pt} 
		\begin{align*}
			\nabla y(u_k) = &-2|c_k|^2 \sum_{i=1}^{L_t-1} \sum_{j=i+1}^{L_t} \left| \left[ \mathbf{B}^H \mathbf{B} \right]_{i,j} \right| \cdot \sin \left( \alpha(u_k) \right) \cdot \frac{2\pi}{\lambda} (\cos \theta_j - \cos \theta_i) \\
			&+2|c_k| \sum_{r=1}^{R} \sum_{i=1}^{L_t} \left| \left[ \mathbf{B} \right]_{r,i} \right| |\boldsymbol{\Lambda}_r| \cdot \sin \left( \beta(u_k) \right) \cdot \frac{2\pi}{\lambda} \cos \theta_i, \tag{16}\label{16}\\
			\nabla^{2} y(u_k) = &-2|c_k|^2 \sum_{i=1}^{L_t-1} \sum_{j=i+1}^{L_t} \left| \left[ \mathbf{B}^H \mathbf{B} \right]_{i,j} \right| \cdot \cos \left( \alpha(u_k) \right) \cdot \left( \frac{2\pi}{\lambda} (\cos \theta_j - \cos \theta_i) \right)^2 \\
			&-2|c_k| \sum_{r=1}^{R} \sum_{i=1}^{L_t} \left| \left[ \mathbf{B} \right]_{r,i} \right| |\boldsymbol{\Lambda}_r| \cdot \cos \left( \beta(u_k) \right) \cdot \left( \frac{2\pi}{\lambda} \cos \theta_i \right)^2. \tag{17}\label{17}
		\end{align*}
	\end{figure*}
	In this section, computer simulation is performed to evaluate the effectiveness of the proposed algorithm. Unless otherwise stated, in the MA-enabled SM-MIMO system, we assume that the distance between the transmitter and receiver is \(d = 40\) m, and each end is equipped with \(T = R = 4\) MAs. We set the minimum distance between MAs as \(D = \lambda/2\), and the wavelength is \(\lambda = 0.05\) m. The upper bound of the moving regions of MAs are set as \(A_t = A_r = 8\lambda\). Using the geometry channel model, assume the numbers of transmit and receive paths are the same, i.e. \(L_{k}^{\mathrm t} = L_{k}^{\mathrm r} \triangleq L\). In this case, the path response matrix is simplified as a diagonal matrix, i.e., \(\mathbf \Sigma = \text{diag} \{\sigma_1, \cdots, \sigma_L\}\) with \(\sigma_\ell\) satisfying \(\sigma_\ell \sim \mathcal CN \left(0,\frac{c^2}{L}\right)\). The signal power gain \(c^2 = C_0 d^{-\alpha}\) is obtained by simplified path loss model, where \(C_0 = -30\) dB is the expected average channel power gain at the reference distance of 1 m and the path loss exponent \(\alpha = 2.5\). AoDs and AoAs are assumed to be independent identically distributed variables, i.e. \( \theta_i \sim \mathcal {U}\left[0, \pi \right] (i = 1, 2, \ldots, L_t) \) and \( \phi_j \sim \mathcal {U}\left[0, \pi \right] (j = 1, 2, \ldots, L_r) \), respectively. The SNR is defined as the ratio of the signal power to the noise power, \(P_T/\sigma^2\).
	\begin{figure*}[!t]
		\centering
		\begin{subfigure}[b]{0.26\linewidth}
        \includegraphics[width=\linewidth]{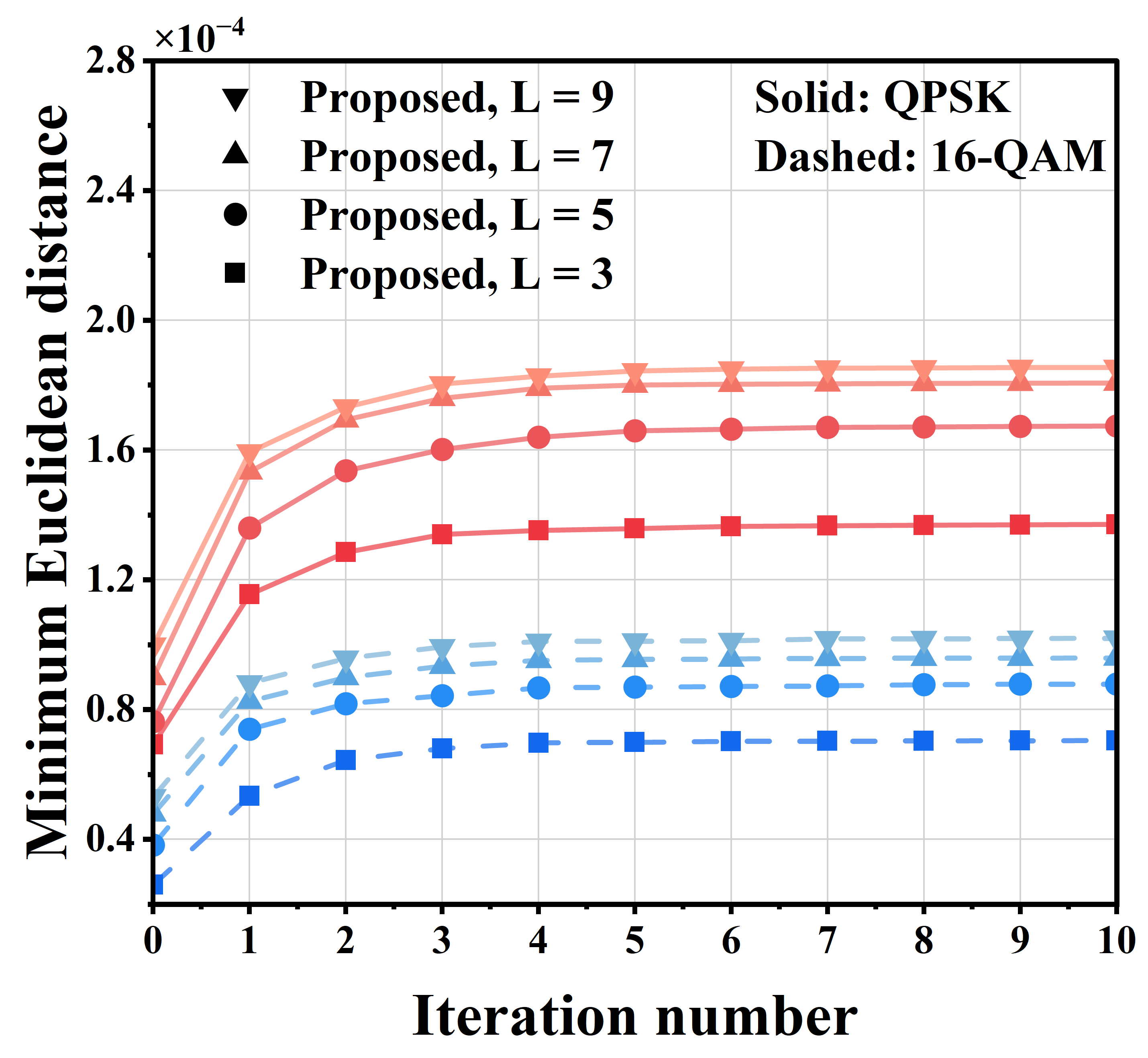}
			\caption{}
			\label{iter}
		\end{subfigure}
		\hfil
		\begin{subfigure}[b]{0.26\linewidth}	\includegraphics[width=\linewidth]{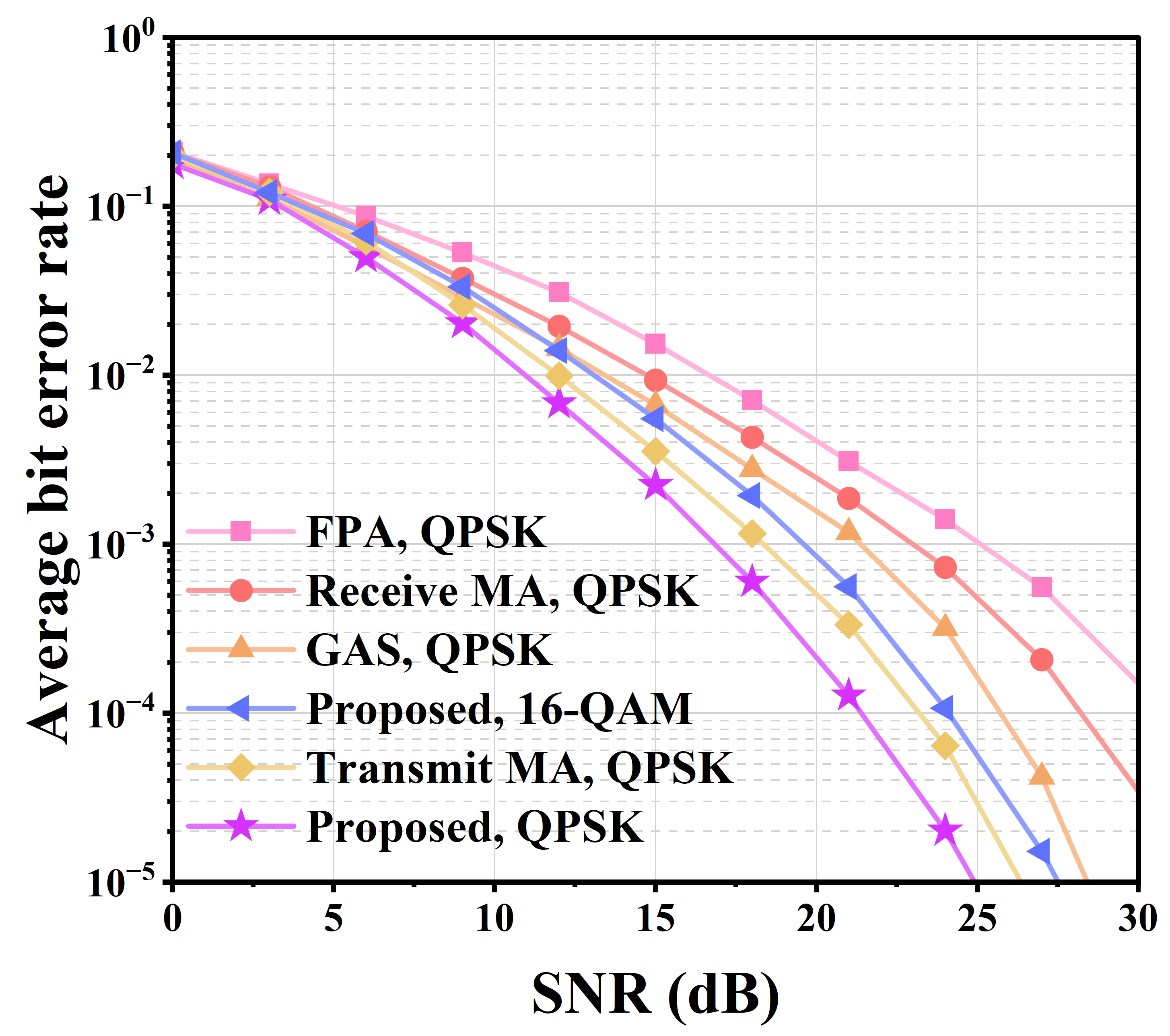}
			\caption{}
			\label{BER1}
		\end{subfigure}
		\hfil
		\begin{subfigure}[b]{0.26\linewidth}
        \includegraphics[width=\linewidth]{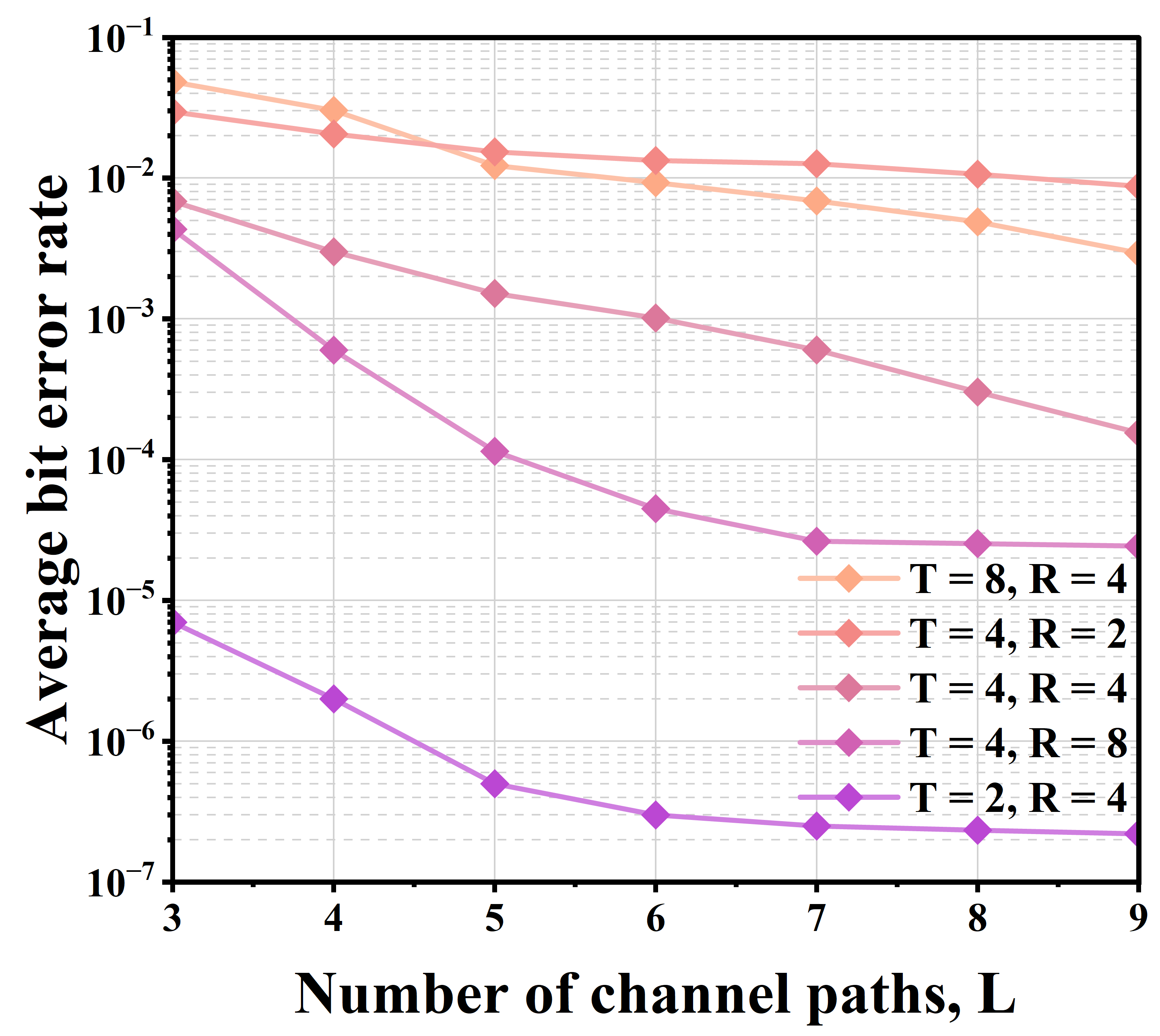}
			\caption{}
			\label{BER2}
		\end{subfigure}
		\caption{Simulation results. (a) Convergence behaviour of Algorithm \ref{Al}. (b) Average BER versus SNR. (c) Average BER versus the number of channel paths.}
        \label{Fig.2}
	\end{figure*}
    
	Fig.~\ref{Fig.2}(\subref{iter}) illustrates the process of maximizing the minimum Euclidean distance, demonstrating the convergence behaviour of the AO algorithm. Notably, the algorithm achieves stability within around five iterations for various modulation schemes, exhibiting excellent convergence performance. Clearly, Increasing the number of channel paths \(L\) can expand the achievable minimum Euclidean distance for each modulation scheme, thereby yielding higher optimal solutions and better BER performance. This is because more channel paths enhance the spatial diversity gain, and therefore improve the overall communication quality.
	
	Fig.~\ref{Fig.2}(\subref{BER1}) compares BER variation with SNR under different antenna configurations. Obviously, the proposed algorithm is superior to other algorithms. At an SNR of 10 dB, the MA algorithm outperforms both the FPA and greedy antenna selection (GAS) algorithms by approximately \(90\%\) and \(50\%\), respectively. Interestingly, the method with MA applied only to the transmitter side is superior to that with MA applied only to the receiver side, indicating that power allocation significantly impacts the current system. The precoder design and antenna position optimization can form a synergistic effect. In addition, for higher order modulation \(M\), the distance between symbols is reduced, which will lead to the system performance decline under the same SNR.
	
	Fig.~\ref{Fig.2}(\subref{BER2}) shows the BER versus the number of channel paths \(L\) when SNR = 12 dB with different antenna configurations. With the increase of the number of channel paths \(L\), the BER of all schemes shows a decreasing trend. This is because the increase of the number of channel paths can bring higher spatial diversity gain.
	\section{Conclusion}
	In this paper, we studied an MA-enhanced spatial modulation system. To improve the error performance, we formulated a BER minimization problem based on the maximum minimum distance criterion through jointly optimizing the transmit beamforming, and the transmit and receive antenna positions. To solve the intractable problem, we proposed an efficient algorithm employing an AO strategy combined with BCD. We showed that the proposed algorithm enjoy a fast convergence rate, and can significantly enhance the BER performance compared with existing benchmarks.
\setcounter{equation}{18}
\begin{appendices}
\section{Convergence of the SCA-Based Algorithm}
Consider the proposed SCA procedure used in the transmit beamforming update in (\ref{11b})--(\ref{12}). For fixed antenna positions $(\boldsymbol u,\boldsymbol v)$, define for each symbol pair $(i,j)$ as
\begin{align}
f_{ij}(\boldsymbol w)\triangleq \|\mathbf{H}\operatorname{diag}\{\boldsymbol w\}(\mathbf{x}_i-\mathbf{x}_j)\|^2= \boldsymbol w^{H}\mathbf A_{ij}\boldsymbol w,    
\end{align}
where $\mathbf A_{ij}=\mathbf X^{H}\mathbf H^{H}\mathbf H\mathbf X\succeq \mathbf 0$ and $\mathbf X=\operatorname{diag}\{(\mathbf x_i-\mathbf x_j)\}$. Since $\mathbf A_{ij}\succeq \mathbf 0$, $f_{ij}(\boldsymbol w)$ is convex in $\boldsymbol w$, and its first-order Taylor expansion at the current point $\boldsymbol w^{n}$ yields
\begin{align}
&\tilde f_{ij}(\boldsymbol w\mid \boldsymbol w^{n})
= f_{ij}(\boldsymbol w^{n}) + 2\Re\{(\boldsymbol w^{n})^{H}\mathbf A_{ij}(\boldsymbol w-\boldsymbol w^{n})\}
=\nonumber\\
&2\Re\{(\boldsymbol w^{n})^{H}\mathbf A_{ij}\boldsymbol w\}-(\boldsymbol w^{n})^{H}\mathbf A_{ij}\boldsymbol w^{n},    
\end{align}
which satisfies $\tilde f_{ij}(\boldsymbol w\mid \boldsymbol w^{n})\le f_{ij}(\boldsymbol w)$ for all $\boldsymbol w$, is tight at $\boldsymbol w=\boldsymbol w^{n}$, i.e., $\tilde f_{ij}(\boldsymbol w^{n}\mid \boldsymbol w^{n})=f_{ij}(\boldsymbol w^{n})$, and is first-order consistent in the sense that $\nabla \tilde f_{ij}(\boldsymbol w^{n}\mid \boldsymbol w^{n})=\nabla f_{ij}(\boldsymbol w^{n})$. According to \cite{marks1978general}, at iteration $n$, the SCA subproblem replaces the original non-convex constraint $f_{ij}(\boldsymbol w)\ge \eta$ by the convex constraint $\tilde f_{ij}(\boldsymbol w\mid \boldsymbol w^{n})\ge \eta$ for all symbol pairs, together with the power constraint $\|\boldsymbol w\|^{2}\le P_{T}$. Owing to the tightness property, the previous iterate $(\eta^{n},\boldsymbol w^{n})$ is feasible for the surrogate problem, and hence the optimal solution $(\eta^{n+1},\boldsymbol w^{n+1})$ satisfies $\eta^{n+1}\ge \eta^{n}$, implying a monotonically nondecreasing objective sequence. Moreover, since $\tilde f_{ij}(\boldsymbol w\mid \boldsymbol w^{n})\le f_{ij}(\boldsymbol w)$, feasibility of the surrogate constraints guarantees feasibility of the original constraints at the new iterate, i.e., $f_{ij}(\boldsymbol w^{n+1})\ge \tilde f_{ij}(\boldsymbol w^{n+1}\mid \boldsymbol w^{n})\ge \eta^{n+1}$. The sequence $\{\eta^{n}\}$ is upper bounded due to the bounded feasible set induced by the transmit power constraint and the bounded antenna movement and spacing constraints, and therefore converges. For the antenna-position update, the surrogate
\begin{align}
y_{\mathrm{lb}}^{n}(u_{k}) = y(u_{k}^{n}) + \nabla y(u_{k}^{n})(u_{k}-u_{k}^{n})-\frac{\epsilon_{k}}{2}(u_{k}-u_{k}^{n})^{2},   
\end{align}
which is constructed such that $\epsilon_{k}\mathbf I-\nabla^{2}y(u_{k})\succeq \mathbf 0$ over the feasible interval, ensuring that $y_{\mathrm{lb}}^{n}(u_{k})$ is concave and satisfies the same properties of global lower-boundedness, tightness, and first-order consistency. Consequently, each antenna update also yields a nondecreasing objective value while preserving feasibility. Since the overall AO--BCD procedure sequentially updates $\boldsymbol w$, $\boldsymbol u$, and $\boldsymbol v$ using such SCA steps, the objective value $\eta$ is monotonically nondecreasing and converges to locally optimal solution.
\end{appendices}

\end{document}